\documentclass[% 
reprint,
 amsmath,amssymb,
 aps,
floatfix,
]{revtex4-2}

\usepackage{graphicx}% Include figure files
\usepackage{dcolumn}% Align table columns on decimal point
\usepackage{bm}% bold math
\usepackage[x11names,dvipsnames]{xcolor}

\begin{document}

%\preprint{APS/123-QED}

\title{The Second Love Number of Dark Compact Planets  \\ and Neutron Stars with Dark Matter}
\author{Yannick Dengler}
 \email{dengler@itp.uni-frankfurt.de}
\author{J\"urgen Schaffner-Bielich}%
 \email{schaffner@astro.uni-frankfurt.de}
\affiliation{%
 Institut f\"ur Theoretische Physik, J. W. Goethe Universit\"at,\\Max von Laue-Str.~1, 60438 Frankfurt am Main, Germany\\
}%


\author{Laura Tolos}
\email{tolos@ice.csic.es}
\affiliation{Institute of Space Sciences (ICE, CSIC), Campus UAB,  Carrer de Can Magrans, 08193 Barcelona, Spain; \\ Institut d'Estudis Espacials de Catalunya (IEEC), 08034 Barcelona, Spain;\\ Faculty  of  Science  and  Technology,  University  of  Stavanger,  4036  Stavanger,  Norway; \\
Frankfurt Institute for Advanced Studies, Ruth-Moufang-Str. 1, 60438 Frankfurt am Main, Germany}%

\date{\today}% It is always \today, today,

\begin{abstract}
%{\color{orange} {(Laura, Juergen, Yannick: to be improved)}}
We study the mass-radius relation and the second Love number of compact objects made of ordinary matter and non-selfannihilating fermionic dark matter for a wide range of dark matter particle masses, and for the cases of weakly and strongly interacting dark matter.
We obtain stable configurations of compact objects with radii smaller than 10 km and masses similar to Earth- or Jupiter-like stellar objects. 
In certain parameter ranges we find second Love numbers which are markedly different compared to those expected for neutron stars without dark matter. Thus, by obtaining the compactness of these compact objects and measuring their tidal deformability from gravitational wave detections from binary neutron star mergers, the extracted value of second Love number would allow to determine the existence of dark matter inside neutron stars irrespective of the equation of state of ordinary matter. 
\end{abstract}

\maketitle

%\tableofcontents

\section{\label{sec:introduction}Introduction}

Astrophysical and cosmological observations indicate that most of the mass of
the universe appears in the form of dark matter (DM) \cite{Planck:2013pxb,SDSS:2014iwm,KHLOPOV2021103824}. The nature of DM is however still elusive. Whereas there are direct methods for detecting DM using particle accelerators \cite{ATLAS:2012ky,CMS:2012ucb} or analyzing DM scattering off nuclear targets in terrestrial detectors  \cite{Klasen:2015uma}, constraints on the properties of DM can be extracted by studying the effects of DM on compact objects, such as white dwarfs and neutron stars.  Indeed, the possible gravitational collapse of a neutron star due to accretion of DM can set bounds on the DM properties \cite{Goldman:1989nd,Kouvaris:2011fi,Fuller:2014rza,Acevedo:2020gro}. Also, constraints on DM can be obtained from stars that accrete DM during their lifetime and then collapse into a compact star, inheriting the accumulated DM  \cite{Kouvaris:2010jy}. Moreover, the cooling process of compact objects can be affected by the capture of DM, which subsequently annihilates \cite{Kouvaris:2007ay,Bertone:2007ae,Kouvaris:2010vv,McCullough:2010ai,deLavallaz:2010wp,Sedrakian:2018kdm,Bhat:2019tnz}. At the same time, self-annihilating DM accreted onto neutron stars may change significantly their kinematical properties  \cite{PerezGarcia:2011hh} or provide a mechanism to seed compact objects with long-lived lumps of strangelets \cite{PerezGarcia:2010ap}. Furthermore, neutron stars that accommodate non-self annihilating DM have emerged as an interesting astrophysical scenario, where to analyze the effects of DM onto hadronic matter (or even quark matter) under extreme conditions \cite{1983SvA....27..371B,1991SvA....35...21K,deLavallaz:2010wp,Li:2012ii,Sandin:2008db,Leung:2011zz,Leung:2012vea,Xiang:2013xwa,Goldman:2013qla,Khlopov:2013ava,Mukhopadhyay:2015xhs,Rezaei:2016zje,Panotopoulos:2017idn,Nelson:2018xtr,Ellis:2018bkr,Gresham:2018rqo,Ivanytskyi:2019wxd,Karkevandi:2021ygv,Sen:2021wev,Guha:2021njn}. In this context, the existence of compact objects with Earth or Jupiter-like masses but unusual small radii have been put forward \cite{Tolos:2015qra,Deliyergiyev:2019vti,DelPopolo:2019nng}, allowing for a new scenario to determine the existence and nature of DM.

Recently, the detection of gravitational waves (GW) from the merger of a binary neutron star system has opened a new venue for probing the existence of DM \cite{LIGOScientific:2017vwq,LIGOScientific:2020aai}. Whereas the structure of neutron stars can be modified by DM in the post-merger phase \cite{Ellis:2017jgp,Bezares:2019jcb,Horowitz:2019aim,Bauswein:2020kor}, the GW signal also depends on the deformation of the binary neutron star system in the inspiral stage. This information is encoded in second Love number and, hence, in the tidal deformability \cite{Hinderer:2007mb,Hinderer:2009ca}. The presence of DM will change the tidal deformability, thus allowing for constraining DM properties. Also, the combination of GW detections with X-ray astronomy, from NICER \cite{Miller:2019cac,FermiGamma-rayBurstMonitorTeam:2020wqj} or eXTP \cite{Watts:2018iom}, and radio, e.g. SKA \cite{Watts:2014tja}, will help to determine the presence and nature of DM in compact objects, such as neutron stars.

%El{\color{blue}lis:2018bkr: tidal
%Nelson:2018xtr: tidal

% references taken from 2109.03801

In this paper we study compact stellar objects that are made of ordinary neutron star matter (OM) admixed with non-self annihilating DM. In order to do so, we have corrected \footnote{A numerical error was found in the computation of Ref.~\cite{Tolos:2015qra} and an erratum paper was written} and extended the previous works \cite{Tolos:2015qra,Tolos:2015qra,Deliyergiyev:2019vti} by analysing the mass and radius configurations of these compact objects for different DM particle masses, and for weakly and strongly interacting DM. 
 Indeed, in the present manuscript we have performed a detailed and deeper analysis of the different configurations, delineating the trends with the mass and the strength of the DM interaction, in particular for the solutions that differ from the neutron-star  and white-dwarf branches. Our final goal is to determine the second Love number of these new mass-radius configurations in view of the recent results coming from GW events. 
The tidal deformability has been investigated for boson stars in Refs.~\cite{Sennett:2017etc,Maselli:2017vfi}, for pure fermionic dark stars \cite{Wahidin:2019juh,Wahidin:2019xcj}, and for a mixture of bosonic DM and neutron star matter in Refs.~\cite{Ellis:2018bkr,Karkevandi:2021ygv}. Note that Ellis et al. in Ref.~\cite{Ellis:2018bkr} mention fermionic DM but the calculations are only done for bosonic DM. There is also one recent work adding fermionic DM in neutron stars to constrain DM parameters from pulsar data \cite{DiGiovanni:2021ejn}. However, a calculation of the Love number is missing. In all the above cases, the parameters of DM are confined so that the compact star configurations have masses of the order of ordinary neutron stars. A systematic and parametric study in terms of different DM masses and interactions has not been done in the literature. To the best of our knowledge, there is no work yet studying the second Love numbers as function of compactness for all different possible configurations of mass and radius in the case of fermionic DM admixed with fermionic OM, as discussed in the present manuscript {\color{blue} \footnote{After the submission of the present manuscript, a  contribution to the proceedings of PANIC2021 \cite{Sagun:2021oml} and a special issue to Galaxy journal \cite{Das:2021hnk} appeared on tidal deformability of fermionic dark matter admixed with neutron star matter.}}.

%Dark matter is one of most researched topic in modern physics, yet, the true nature is still almost entirely unknown. The evidence for dark matter is striking. Groups of scientist try to gather information about possible dark particles in collider experiments or via direct observation. Some try to link astronomical measurements to a dark particle.  In this work, an inspiral including a neutron star is used as an astronomical probe for dark matter. The neutron star is assumed to inhabit dark matter either as a dark core or as a dark halo surrounding the star. A neutron star is described by its equation of state (EoS). The EoS neutron star matter is approximately known up to the saturation density of nuclear matter. But it is likely that neutron stars reach much higher densities in the core. Many scientists expect quark matter in the core of neutron stars. Such stars are called hybrid-stars. From the EoS, one obtains the mass and the radius of a star with the TOV-equation. Therefore, the EoS is constraint by the most accurate measurements of masses and radii of neutron stars. 

\section{Formalism}

\subsection{TOV-equations \\ for dark compact stellar objects}

In the following, we investigate compact objects that are made of OM admixed with non-self-annihilating DM, following the steps of Ref.~\cite{Tolos:2015qra}. These two types of matter are described by two fluids that only interact gravitationally \cite{Sandin:2008db}.  One proceeds by solving simultaneously the coupled TOV equations for OM and DM in dimensionless form as
\begin{eqnarray}
&&\frac{dp'_{OM}}{dr}=-(p'_{OM}+\varepsilon'_{OM})\frac{d \nu}{dr}, \nonumber \\
&&\frac{dm_{OM}}{dr}=4 \pi r^2 \varepsilon'_{OM}, \nonumber \\
&&\frac{dp'_{DM}}{dr}=-(p'_{DM}+\varepsilon'_{DM}) \frac{d \nu}{dr}, \nonumber \\
&&\frac{dm_{DM}}{dr}=4 \pi r^2 \varepsilon'_{DM}, \nonumber \\
&&\frac{d \nu}{dr}=\frac{(m_{OM}+m_{DM}) + 4 \pi r^3(p'_{OM}+p'_{DM})}{r(r-2(m_{OM}+m_{DM}))}, 
\label{eq:tovs}
\end{eqnarray}
where $p^\prime$ and $\varepsilon^\prime$ are the dimensionless pressure and energy density respectively, defined as $p^\prime = \frac{p}{m_f^4}$ and $\varepsilon^\prime = \frac{\varepsilon}{m_f^4}$, with $m_f$ being either the neutron mass ($m_n$) or the mass of the fermionic DM particle ($m_F$). We choose the latter one. Then, the physical mass and radius for each specie are $R=
 (M_p/m_F^2) \, r$ and $M= (M_p^3/m_F^2) \,  m$, respectively, where $M_p$ is the Planck mass \cite{Narain:2006kx}.

The EoS for OM is given by the equation-of-state EoSI
from \cite{Kurkela:2014vha}. The different EoSs obtained in \cite{Kurkela:2014vha} are
constrained by using input from low-energy nuclear physics using chiral effective theory and the high-density
limit from perturbative QCD. The regime between these two limits is described by interpolated piecewise polytropes that are restricted by observational pulsar data. In particular, the EoSI
is the most compact one with maximum masses of 2${\rm M}_{\odot}$. We, moreover,
map EoSI with an inner and outer crust EoS using \cite{Inner_crust_EoSI} and
\cite{Outer_crust_EoSI}, respectively. For $\rho < 3.3 \times 10^{3}$
$\rm{g/cm^3}$ we use the Harrison-Wheeler EoS \cite{Harrison_Wheeler_Book}.
%{(\color{red}\\ Why are we using a curved R? Should`nt it be a density?)}

In the case of DM, we consider non-selfannihilating fermions. Two cases are studied, weakly and strongly interacting DM. The strength of the interaction is controlled by the strength parameter, $y=m_F/m_I$, which is defined as the ratio between the  mass of the fermionic dark particle $m_F$ and the interaction mass scale $m_I$ (see Eqs.~(34-35) in Ref.~\cite{Narain:2006kx}). For strong interactions, $m_I \sim 100~{\rm MeV}$  (interaction scale related to the exchange of vector mesons) while for weak interactions $m_I \sim 300~{\rm GeV}$ (exchange of W and Z bosons). We consider two extreme cases, that is, $y=10^{-1}$ and $y=10^{3}$ for weakly and strongly interacting DM, respectively.

%The interaction is described by an interaction term for the pressure and the energy density that is proportional to the number density n squared. $P_{int}=\varepsilon_{int}=y^2 n^2$. y is a scale for the strength of the interaction, that can be motivated with an interaction scale $m_I$($y=\frac{m_F}{m_I}$). $m_I$ is the mass of a hypothetical interaction particle, which would be $m\approx m_\pi$ for a strong interaction of $m_I \approx m_{W^\pm,Z^0}$ for a weak interaction. Besides the strength of the interaction the mass of the dark fermion can be varied. A mass of $m_F = m_N$ corresponds to mirror dark matter ($m_N$ is the mass of a nucleon). For the case of asymmetric DM, the fermion mass would be $m_F = 5 m_N$. The 100 GeV case is a approximation for a hypothetical neutralino.

%As the two types of matter are described by two separate fluids, the pressure may differ at the same point in space. Especially the central pressure may differ. Therefore one obtains unique solutions to the TOV-equation for every choice of the ratio of central pressure.

%\begin{equation}
%r = \frac{P_{0,DM}}{P_{0,OM}}
%\end{equation}

In order to solve the TOVs one needs to determine the initial central pressure for both species. Following Ref.~\cite{Tolos:2015qra} we fix the ratio of central pressures to different values to take into account different scenarios regarding the impact of the DM component. In particular, in this paper we consider
\begin{equation}
\mathcal{R} = \frac{p_{0,DM}}{p_{0,OM}}\left(\frac{m_n}{m_{F}}\right)^4 ,
\label{eq:ratio}
\end{equation}
where $p_0$ is the central pressure for DM or OM.
%\\{(\color{red} In the end, the ratio r is chosen arbitrarily. The values for r that yield the most interesting results are chosen. I just mentioned in my thesis that one should use higher values for r for higher dark fermion masses! Btw: I just noticed a typo in my master thesis where I wrote: r=$p_{OM} / p_{DM}$)} Given the possible values of the mass of the dark particle, we vary $m_{F_{DM}}$ from 1 to 100 GeV.

Moreover, for the analysis of the compact objects with DM content, one first has to perform an
analysis of the stable configuration for both OM and DM.  The stability arguments can be found, for example, in \cite{Stability_textbook}, where the
stability of the different radial modes in a star is analyzed. In the following we give a short description.

%As already mentioned, unstable solutions to the TOV-equation occur for certain central pressures. An unstable solution diverges, when a small radial perturbation is applied to the star. 

In order to check the stability of a given configuration, one has to consider small radial perturbations of the equilibrium configuration by solving the Sturm-Liouville eigenvalue equation, that yields eigenfrequencies $\omega_n$ \cite{Stability_textbook}. The eigenfrequencies of the different modes form a discrete hierarchy $\omega_n^2 < \omega_{n+1}^2$ with n=0,1,2..., being real numbers. A negative value of $\omega_n^2$ leads to an exponential growth of the radial perturbation and collapse of the star. The determination of the sign of the mode results from the analysis of the mass of the star versus the mass density or radius. When the mass reaches an extremum, a mode changes sign. What mode changes sign depends whether the mass-radius relation is going clock- or counterclockwise at the extremum. A clockwise orientation leads to a eigenfrequency going positive again, that is, a mode becoming stable again. A counterclockwise orientation leads to an additional unstable mode. Thus, starting at low mass densities where all eigenfrequencies are positive, one can perform the stability analysis for higher mass densities studying the change of sign of the different modes while keeping the hierarchy among them. Only when all eigenfrequencies are positive, the star will be stable \cite{BTM_1966,Alford_2017}. In this way, one can study the simultaneous stable regions for OM and DM, as done in Ref.~\cite{Tolos:2015qra}. 

We close this discussion on the stability of two-fluid configurations in general relativity with a word of caution.
While our prescription is valid for one-fluid configurations, see Ref.~\cite{BTM_1966},
a full stability analysis for two-fluids would require to solve the coupled Sturm-Liouville eigenvalue equations. 
More rigorous investigations in this direction just appeared recently involving time-consuming numerical computations \cite{Kain:2020zjs,Kain:2021hpk,Jimenez:2021nmr} (for earlier work see \cite{Leung:2011zz,Leung:2012vea}). 
For neutron stars with an admixture of non-interacting DM stable solutions were found in parameter regions which did not seem feasible in a naive analysis \cite{Kain:2020zjs,Kain:2021hpk}.
The stability of quark stars admixed with DM was studied by solving the combined Sturm-Liouville eigenvalue equations for the two-fluid TOV equations in Ref.~\cite{Jimenez:2021nmr}.
Their results shows that only small quark matter masses are dynamically stable leading to stable dark strange planets and dark strangelets, and that the stability of DM stars are mainly affected for small fermion masses. 
We leave a full stability analysis on the stability of the OM and DM configurations studied here as a topic of investigation for future numerical studies.

%An imaginary $\omega_n$ leads to an exponential growth and the collapse of the star. There is a trick to check, whether a solution is stable or not. For this, one starts at a solution that is known to be stable (This is the white dwarf branch in our case) and increases the central pressure. When the mass reaches an extremum, i.e. $\frac{dM}{dP_0}=0$, a mode changes sign. What mode changes sign depends on, whether the M-R-R is going clock- or counterclockwise at the extremum. A clockwise orientation leads to a mode going stable again. A counterclockwise orientation leads to an additional unstable mode. Only when all modes are stable, the solution of the TOV-equation is stable. For two fluids, as investigated in this paper, this stability analysis has to be performed twice. Only if the ordinary and dark matter is stable, the entire star is stable.

\subsection{Tidal Deformability \\ for dark compact stellar objects}

The detection of gravitational waves of binary neutron star mergers, in particular from the GW170817 event \cite{LIGOScientific:2017vwq}, has posed recent constraints of the EoS through the so-called tidal deformability during the inspiral phase.

The tidal deformability $\lambda$ measures the induced
quadrupole moment, $Q_{ij}$, of a star in response to the tidal field of the
companion, $\mathcal{E}_{ij}$ \cite{Hinderer:2007mb,Hinderer:2009ca}
\begin{equation}
Q_{ij}=-\lambda \mathcal{E}_{ij} .
\end{equation}

%The tidal deformability of an object is defined for objects inside an external tidal field. The tidal deformability is defined as the proportionality between the deformation of the object and the strength of the tidal field. 
%\begin{equation}
%    Q_{ij} = \lambda \epsilon_{ij}
%\end{equation}

The tidal deformability is connected to the dimensionless second Love number $k_2$ as
\begin{equation}
    \lambda=\frac{2}{3}k_2 R^5 ,
\end{equation}
where $R$ is the radius of the star. The tidal Love number can be calculated from
\begin{eqnarray}
k_2&=&\frac{8C^5}{5} (1-2C)^2 [2+2C(y_R-1)-y_R] \times \nonumber \\
&& \{ 2C[6-3y_R+3C(5y_R-8)]+\nonumber \\
&&4C^3[13-11y_R+C(3y_R-2)+ 2C^2(1+y_R)] + \nonumber \\
&&3 (1-2C)^2[2-y_R+2C(y_R-1)] {\rm ln}(1-2C) \}^{-1}  , \ \ \ \ \ \
\end{eqnarray}
with $C$ being the compactness parameter, that for the case of dark compact objects is given by $C=M/R$, with $M=M_{DM}+M_{OM}$ and $R=R_{\rm max} (R_{OM}, R_{DM})$. The quantity
$y_R\equiv y(R)$ is obtained by solving, together with the TOVs of Eq.~(\ref{eq:tovs}),  the following differential equation for $y$ \cite{Postnikov:2010yn}
\begin{equation}
    r\frac{dy(r)}{dr}+y^2(r)+y(r)F(r)+r^2Q(r)=0 ,
\end{equation}
with
\begin{equation}
    F(r)=\frac{r-4\pi r^3\left((\varepsilon'_{OM}+\varepsilon'_{DM})-(p'_{OM}+p'_{DM})\right)}{r-2(m_{OM}+m_{DM})} ,
\end{equation}
\begin{eqnarray}
    Q(r)&=&\frac{4\pi r}{r-2(m_{OM}+m_{DM})}\nonumber \\
    &\times& \left[5(\varepsilon'_{OM}+\varepsilon'_{DM})+9(p'_{OM}+p'_{DM})\right.\nonumber \\
    &+&\left.\frac{\varepsilon'_{OM}+p'_{OM}}{c_{s,OM}^2}+\frac{\varepsilon'_{DM}+p'_{DM}}{c_{s,DM}^2}-\frac{6}{4\pi r^2}\right] \nonumber \\
    &-&4\left[\frac{(m_{OM}+m_{DM})+4\pi r^3(p'_{OM}+p'_{DM})}{r^2(1-\frac{2(m_{OM}+m_{DM})}{r})}\right]^2 ,
\end{eqnarray}
where $c_s(r)^2=dp'/d\varepsilon'$ is the squared speed of sound for OM ($c_{s,OM}$) and DM ($c_{s,DM}$). The starting condition for $y$ in the center of the star is $y(r=0) = 2$. 
Once $k_2$ is known, the dimensionless tidal deformability $\Lambda$ can then
be determined by the relation
\begin{equation}
\Lambda=\frac{2k_2}{3C^5} .
\label{eq:k2-lambda}
\end{equation}

\section{Results}
In this section we show our results for the mass-radius relation and second Love number of compact stellar objects with DM content.

\begin{figure*}
\includegraphics[width=.8\textwidth]{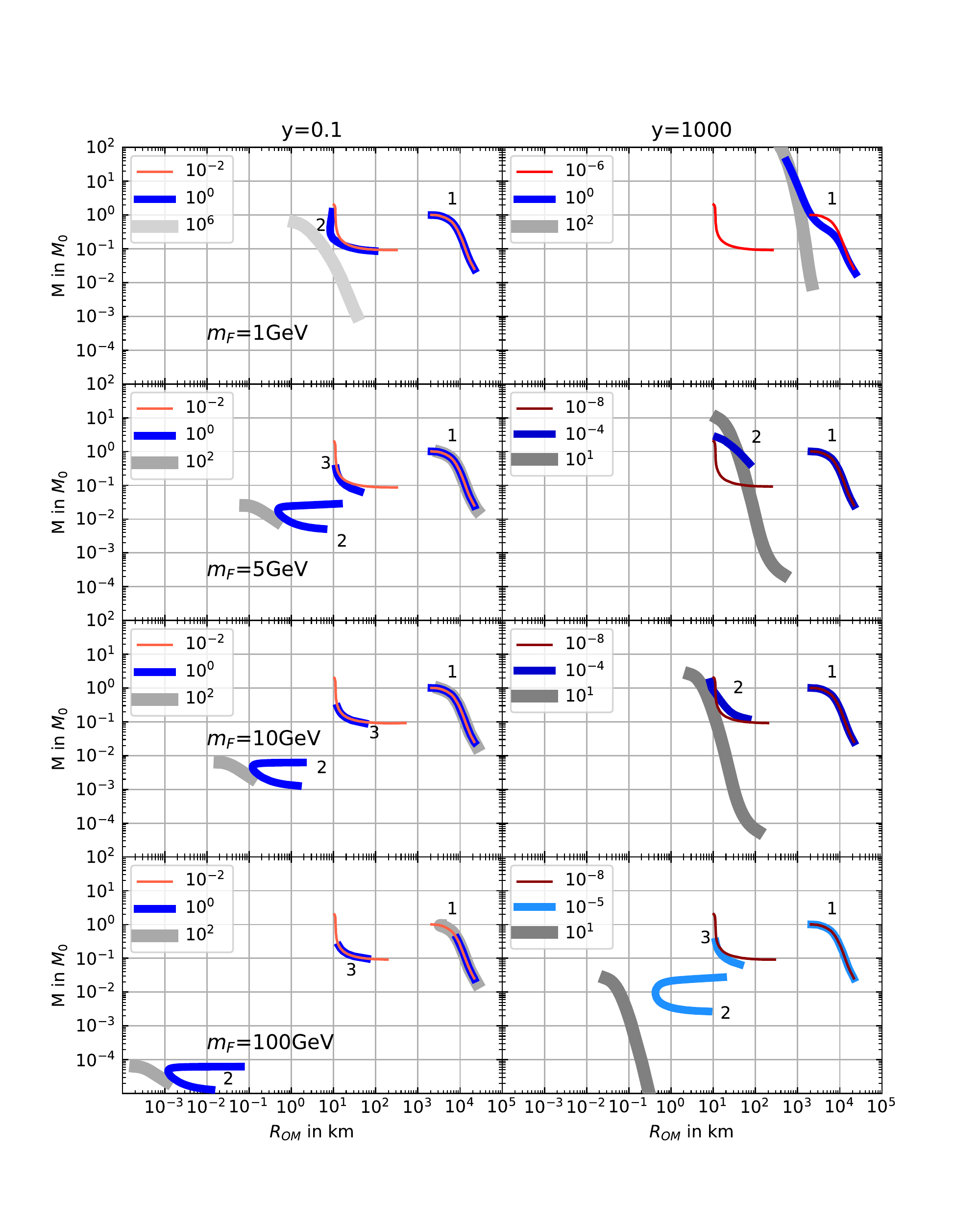}% Here is how to import EPS art
\caption{\label{fig:M_R} The total mass as a function of the radius of ordinary matter for compact stellar objects using different DM particle masses (different rows), and for weakly interacting DM (left column) and strongly interacting one (right column).}
\end{figure*}

%\subsection{Mass and Radius of Compact Objects}
Figure~\ref{fig:M_R} shows the stable solutions for compact stellar objects with DM content. We show the total mass of the compact object as a function of the radii of OM (visible matter) for different masses of the dark particle (different rows) and two extreme interaction strengths for the DM component, that is, weakly interacting DM $y=0.1$ (left column) and strongly interacting DM $y=1000$ (right column). In each panel we show the masses and radii for three different ratios between the central pressure of DM with respect to the OM one ($\mathcal{R}$ in Eq.~(\ref{eq:ratio})) with red (thin), blue (medium) and grey (thick) lines, respectively. Depending on this ratio, one, two or even three stable branches might appear.  Note that we numbered the stable branches with '1', '2' or '3' for the middle value of the ratio (blue lines). Also, note that the present calculation corrects and extends the computation done in Refs.~\cite{Tolos:2015qra,Deliyergiyev:2019vti} for a larger set of DM particle masses.

%Every stable branch is numerated. Higher numbers are results at higher central pressures. 

For low values of the ratio $\mathcal{R}$ (red lines), the DM content is negligible. Therefore, two stable branches are clearly visible resulting from the dominance of the OM component, that is, the one associated to white dwarfs that appears for radii  larger than $10^{3}$ km with maximum masses of $1M_\odot$ and the neutron star branch for radii $10-10^2$ km and maximum masses of $2M_\odot$. 

As we increase the ratio $\mathcal{R}$ up to  $10^2$ (grey lines), whereas we still obtain the white-dwarf branch for the weakly interacting case, for higher central pressures, a new branch appears which is dominated by the DM component. The masses and radii of this branch depend on the mass of the DM particle and the interaction strength.  The total mass is dominated by the mass of the DM component, scaling with the inverse of the squared of the dark particle mass, as discussed in Ref.~\cite{Narain:2006kx}.  Moreover, as seen in Ref.~\cite{Narain:2006kx}, for the weakly interacting case, the slope of the mass-radius curve for masses well below the maximum mass is proportional to $R_{DM}^{-3}$. For strongly interacting DM, the mass-radius relation is constant for a given $R_{DM}$, as presented in Ref.~\cite{Narain:2006kx}. Note that this behaviour can not be seen in Fig.~\ref{fig:M_R} as the total mass of the star is plotted against the radius of the OM and not the radius of DM.

%\begin{table}[]
%\begin{tabular}{|l|l|l|l|l|}
%\hline
%Fermion Mass in GeV & 1    & 5      & 10 & 100                      \\ \hline
%weak interaction    & 21   & 83 (8) & 99 & $\approx 100$ \\ \hline
%strong interaction  & 98.5 & 90     & 71 & 81                 \\ \hline
%\end{tabular}
%\caption{\label{table:amounts} Proportion of dark matter mass in \% of the total mass of the star at the maximum mass configuration for the second stable branch in the mixed case. The white dwarf branch is not considered. }
%\end{table}

\begin{table}
\begin{tabular}{|l|l|l|}
\hline
$m_F[{\rm GeV}]$ & WI DM (\%) & SI DM (\%) \\ \hline
1 & 21 & 98.5 \\ \hline
5 & 81 (8) & 90 \\ \hline
10 & 99 (2) & 71 \\ \hline
100 & $\sim$ 100 ($\sim 0$) & 81 (8) \\ \hline
\end{tabular}
\caption{\label{table:amounts} Amount of DM mass ($M_{DM}$) with respect to the total mass ($M$) in \% at the maximum mass-radius configuration for the '2' stable branch (labelled with '2' in Fig.~\ref{fig:M_R}) for different values of the DM particle ($m_F$) and for the case of intermediate values of the ratio of pressures between the DM and OM components. We consider weakly-interacting DM (WI DM) and strongly interacting DM (SI DM). Note that for the strongly interacting case with $m_F=1$ GeV there is only one stable branch  with '1'. The number in the brackets corresponds to the amount of DM mass for the '3' stable branch (if it exists), labelled with '3' in Fig.~\ref{fig:M_R}. }
\end{table}

We now turn our attention to intermediate values of the ratio $\mathcal{R}$ (blue lines) in  Fig.~\ref{fig:M_R}. For that case, we labelled the different curves in Fig.~\ref{fig:M_R} with '1', '2' or even '3', being '1' the curve closer to the white-dwarf branch for low values of $\mathcal{R}$.  Moreover, we show in Table~\ref{table:amounts} the amount of DM mass with respect to the total mass (in percent) at the maximum mass for '2' in Fig.~\ref{fig:M_R}. Note that for the strongly interacting case with $m_F=1$ GeV, there is only one stable branch  with '1'. We consider weakly interacting DM (WI DM) and strongly interacting DM (SI DM).  The number in the brackets corresponds to the amount of DM mass at the maximum mass for the '3' stable branch (in case that it exists), labelled with '3' in Fig.~\ref{fig:M_R}. 

We start by analysing the weakly interacting DM and intermediate values of $\mathcal{R}$. For that case and DM particle mass of 1 GeV (close to the nucleon mass), the mass-radius relation '2' for $\mathcal{R}=1$ is close to the neutron-star branch of $\mathcal{R}=10^{-2}$, being slightly shifted to lower masses and radii. The amount of DM mass with respect to the total mass at the maximum mass-radius configuration  is 21\%, as shown in Table~\ref{table:amounts}. When the mass of the DM particle is increased, the '2' branch drops below the neutron-star branch  (observed for low values of $\mathcal{R}$), and the mass-radius stable configurations appear closer to the DM dominated branch (seen for large values of $\mathcal{R}$) and away from the neutron and white-dwarf branches (obtained for low values of $\mathcal{R}$). The mass and radius of these solutions, already described as dark compact planets (DCPs) in Ref.~\cite{Tolos:2015qra}, scale with the inverse of the DM mass squared. Note that the larger the DM mass is, the larger the central pressure of DM is needed in order to obtain the DCPs (see Eq.~(\ref{eq:ratio})). Moreover, the amount of DM for the '2' stable branch increases with increasing DM particle mass, getting closer to the DM dominated branch (observed for large values of $\mathcal{R}$), whereas the '3' branch (when it exists) lies closer to the neutron-star branch (seen for small values of $\mathcal{R}$) and, hence, contains a small amount of DM.

For the strongly interacting case and intermediate values of $\mathcal{R}$, the masses and radii of the '2' branch are larger compared to neutron-star branch (seen for low values of $\mathcal{R}$). These solutions for different DM masses contain large amounts of DM, as seen in Table~\ref{table:amounts}. Note that for the DM mass of 10 GeV, we have a neutron star inside a larger heavy DM halo. Only for a DM mass of 100 GeV, the '2' branch drops below the neutron-star branch (observed for low values of $\mathcal{R}$).  Moreover, for a DM mass of 100 GeV, a branch labelled with '3' appears close to the neutron-star branch (obtained for low values of $\mathcal{R}$), repeating the pattern seen for the weakly interacting case.  This solution contains a small fraction of DM at the maximum mass, as also seen for the '3' branch in the weakly interacting case.

%Whereas for a DM mass of 1 GeV, this solution reaches masses larger than 20 $M_\odot$ {(\color{blue} Yannick: where is curve 2?? or you are talking about curve 1??),  for DM mass of 5 GeV, the masses and radii are above the neutron-star . The star is still mostly made of dark matter. For 10 GeV, the branch looks very similar to the neutron star branch. This configuration is an ordinary light neutron star inside of a larger heavy dark matter halo. If the radius of the dark matter halo has no impact on the radius measurement of the neutron star, this result would look very similar to a neutron star. \newline

%In general, the addition of a small matter of a second kind that only interacts gravitationally to a star shifts the masses and radii to lower values.

\begin{figure*}
\includegraphics[width=.7\textwidth]{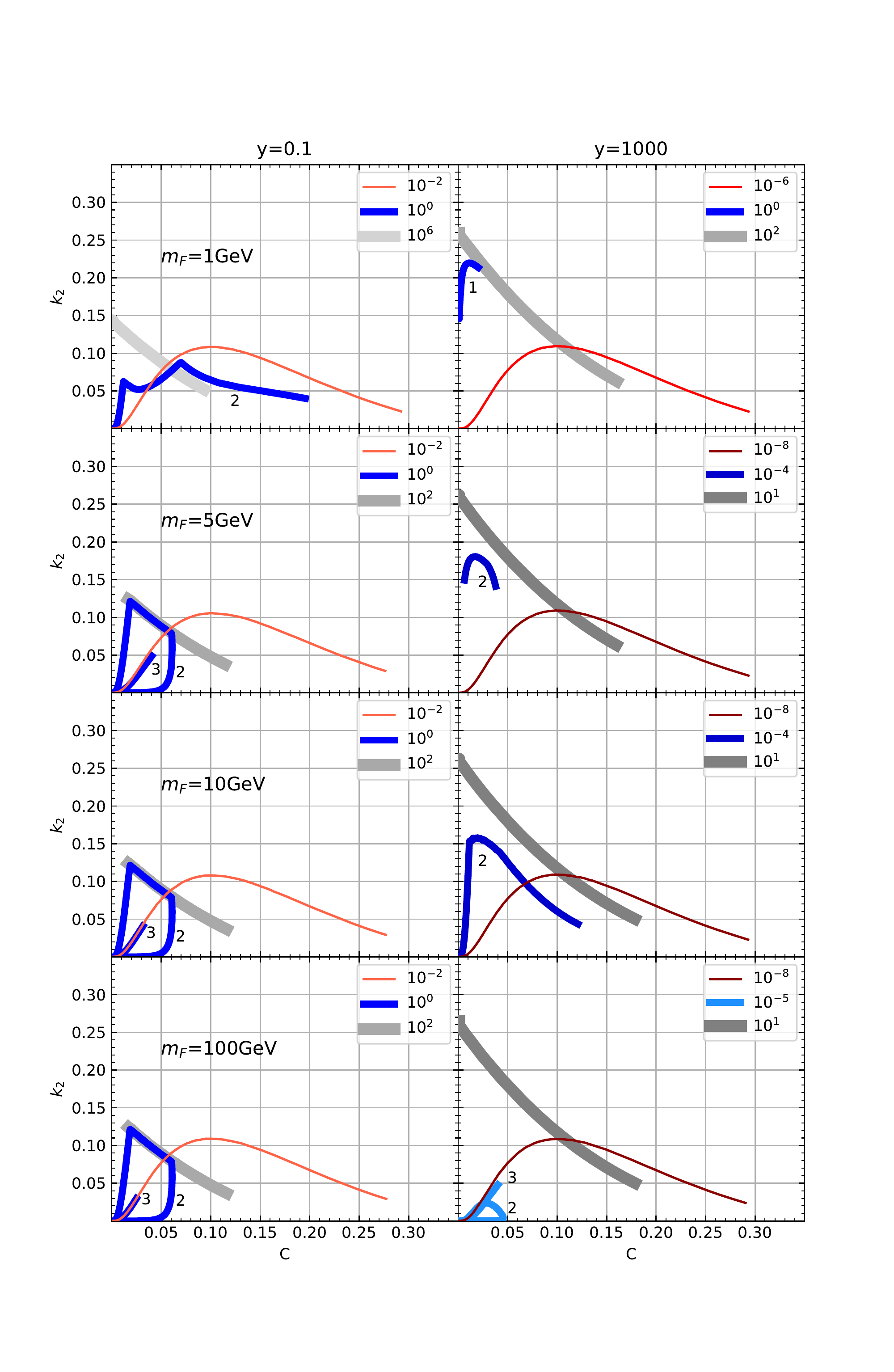}% Here is how to import EPS art
\caption{\label{fig:k_2} The second Love number $k_2$ as a function of compactness $C$ for compact stellar objects using different DM particle masses (different rows) and for weakly interacting DM (left column) and strongly interacing one (right column).}
\end{figure*}

%The Tidal Deformability is closely related to the second Love number $k_2$. \Ref{fig:k_2} shows $k_2$ plotted against the compactness of the star ($C=\frac{M}{R}$). Note that the white dwarf branch is not shown in the plots, as the compactness is $C\sim 10^{-4}$. The curve for an ordinary matter dominated star is close to the results for a realistic equation of state (\cite{Hinderer_2010}). The DM dominated curves are given by the results for a pure Fermi-gas. The compactness for the case without interaction is limited by C=0.11. For the interaction dominated case, it is limited by C=0.21. \\
%An addition of DM to the an ordinary matter star or vice versa lowers the values of $k_2$. For the weakly interacting case with a fermion mass of 1 GeV, the Love number is shifted to lower values. This configuration yields results close to realistic neutron stars. The additional DM has a significant impact on the Love number. \\
%Other configurations do not yield results close to neutron stars. The low interacting DCPs have the same values. One exception is the additional '3' branch for a fermion mass of 5 GeV. As the MRR of the configuration is close to a neutron star, $k_2$ is also comparable.\newline
%Although the branches are close in the MRR, the results may differ in the $k_2$-plots. This is because the Love number is sensitive to the internal structure of the compact object. Further, the compactness is calculated with the larger radius which is the radius of dark matter in the case of a dark halo surrounding an ordinary matter core.

Next, we analyze the second Love-number $k_2$ of the different star configurations shown in Fig.~\ref{fig:M_R}. As seen in Eq.~(\ref{eq:k2-lambda}), a large second Love-number means that the star is deformed easily by an external tidal field (large values of $\Lambda$).

Fig.~\ref{fig:k_2} shows $k_2$ against the compactness of the star $C$ for the same star configurations displayed in Fig.~\ref{fig:M_R}, that is, for different DM masses (different rows) and for weakly interacting DM, $y=10^{-1}$ (left column) and strongly interacting DM, $y=10^3$ (right column). Note that the white dwarf branch is not visible in the plots for all cases because the compactness is too low ($C\approx10^{-4}$). 

We start by considering low values of $\mathcal{R}$ for all masses and both strengths (red lines). Those are associated to the neutron-star branch, that is, configurations dominated by OM with only a small amount of DM. The values of $k_2$ follow the expected pattern for neutron stars with an hadronic core, as described in Ref.~\cite{Hinderer:2009ca}. By increasing the central pressure, one moves from $k_2 \rightarrow 0$ at $C \rightarrow 0$ to larger values of $k_2$. At a compactness of $C\approx0.1$, $k_2$ reaches the maximum value and then decreases for large compactness. The maximum compactness of $\approx0.29$ is achieved by neutron stars with $R\sim 10$ km with a mass of 2 $M_\odot$.

Then we study large values of $\mathcal{R}$ for all masses and both strengths (grey lines). As expected, the mass-radius configurations are close to the purely fermionic case \cite{Narain:2006kx}. The larger the central pressure is, the smaller the $k_2$ is with increasing compactness. Moreover, for strongly interacting DM (right column) larger values for $k_2$ and $C$ are achieved as compared to the weakly interacting case. Note that the compactness reached for DM dominated stars should be smaller than the limits dictated for pure fermionic matter. Without interactions, this limit is given by $C=0.11$, whereas for strongly interacting matter it results in $C=0.21$ \cite{Schaffner-Bielich:2020psc}.

We finally turn to intermediate values of the ratio $\mathcal{R}$ (blue lines).  Starting with the weakly interacting case and a DM particle mass of 1 GeV, one notices that the curve follows the trend of the neutron star branch (red line for small values of $\mathcal{R}$), as one would expect by the behaviour of branch '2' in the upper left plot of Fig.~\ref{fig:M_R}. However, starting from a fast rise of $k_2$ with compactness, two kinks occur at $C\approx0.012$ and $C\approx0.07$. Those kinks result from the fact that for the computation of $k_2$ one takes the largest radius of a given mass-radius configuration. The largest radius is $R_{OM}$ for small compactness up to the first kink, where $R_{DM}$ becomes the largest, and back to $R_{OM}$ being the largest in the second kink. After the second kink, the values of $k_2$ are smaller compared to the OM dominated case. The maximum compactness in this configuration is given by $C\approx0.2$. 

When the DM particle mass is increased to 5 GeV for weakly interacting DM, the  behaviour of the $k_2$ for intermediate values of the ratio $\mathcal{R}$ (blue lines) changes drastically as compared to the case of a DM particle mass of 1 GeV. This can be understood by analysing the behaviour of the branches '2' and '3' as compared to the OM dominated case (low values of $\mathcal{R}$) or the DM dominated one (large values of $\mathcal{R}$) in Fig.~\ref{fig:M_R}.  With increasing central pressure, $k_2$ rises steeply until it reaches the DM dominated mass-radius configuration, displayed by the grey line. A kink occurs as the largest radius moves from $R_{OM}$ to $R_{DM}$. Afterwards, the line follows the DM dominated mass-radius configuration and returns to $C \rightarrow 0$. Then, the  branch '3' appears, being close to the OM dominated mass-radius configuration. This pattern repeats for larger DM particle masses. Note that the values for $k_2$ and $C$ are similar for $m_F=5, 10$ and $100$ GeV as both quantities are dimensionless and do not scale with the DM particle mass.  

As for the strongly interacting DM matter, the behaviour of the mass-radius configuration for intermediate values of  $\mathcal{R}$ changes with increasing DM particle mass, as expected from the behaviour of the mass-radius relation in Fig.~\ref{fig:M_R}.  For $m_F=1$ GeV  branches '1' and '2' are connected without an unstable regime in between, reaching values for the mass larger than 20 $M_\odot$ at large radii. As a result, $k_2$ reaches $\approx0.22$, close to the DM dominated branch for large values of $\mathcal{R}$ (grey line). The maximum compactness is then small, $ C\approx0.02$. For $m_F=5$ GeV, $k_2$ first increases up to $C\approx0.18$ and decreases afterwards. The maximum compactness is reached at $ C\approx0.04$. Note that the values of compactness cannot be extracted directly from Fig.~\ref{fig:M_R}, as the OM radius (or visible) radius is plotted there, regardless of the fact that in some mass-radius configuration the largest radius is the DM one. For $m_F=10$ GeV, $k_2$ first increases steeply up to $C\approx0.16$ and, then, it decreases following the pattern of the DM dominated case up to a value of $k_2 \approx0.04$ at a compactness of $C\approx0.12$. Finally, for $m_F=100$ GeV, the branches '2' and '3' for the strongly interacting DM display a similar behaviour as for the weakly interacting DM case in Fig.~\ref{fig:M_R}, although
the mass-radius configuration of the DM dominated case (grey line for large values of $\mathcal{R}$) is well below the branches '2' and '3'. Thus, a similar behaviour for $k_2$ in the weakly and strongly interacting DM cases is expected, in spite of being the values of $k_2$ smaller than those in the DM dominated case. The $k_2$ in branch '2' increases up $k_2\approx0.03$. Afterwards it decreases and returns to zero for $C\approx0.05$. The branch '3' yields similar values as the OM dominated case and reaches a compactness of $C\approx0.04$.

Our results for the neutron-star-like configurations can be compared with previous work on the tidal deformability for neutron stars with an admixture of DM \cite{Nelson:2018xtr,Ellis:2018bkr}. In Ref.~\cite{Nelson:2018xtr} it was found that interacting DM fermions in the MeV-GeV mass range will form a DM halo around neutron stars increasing the tidal deformability. We see such an increase in the radius and therefore in the tidal deformability $\Lambda$ for the strongly interacting case and particularly pronounced for low fermion masses. In Ref.~\cite{Ellis:2018bkr} the presence of a  DM core was leading to a reduced tidal deformability for neutron-star-like configurations for weakly interacting bosons.
We find similar configurations with a reduced radius for the weakly interacting case and light fermion masses 
which would give a reduced tidal deformability $\Lambda$ (see Eq.~\ref{eq:k2-lambda}). 

As a final remark, we should comment on the different possibilities of formation of
compact objects with DM content, that is, by means of the accretion mechanisms of DM onto neutron stars and white dwarfs as well as by the primordial formation of DM clumps surrounded by OM. As argued in Ref.~\cite{Tolos:2015qra} the accretion mechanism (taking into account the local standard DM density) might not explain the quantity of DM obtained inside these compact objects,  whereas a larger DM component is expected in the second case. Moreover, there exists the possibility of having dark compact configurations coming from structure formation from DM perturbations growing from primordial overdensities, as shown in Ref.~\cite{Chang:2018bgx}. Initial density perturbations can produce compact objects of DM with masses from substellar masses to masses of several million solar masses. The key input are vector boson interactions of fermionic DM, which allows for creating out of the initial primordial density perturbations compact objects of various mass scales. The larger the interaction, the smaller the compact objects can be. E.g. compact objects of planetary size can be produced for large interaction strengths or by fragmentation or mergers of compact objects. We note that the interaction used in Ref.~\cite{Hinderer:2007mb} via vector boson exchange gives the kind of EoS for DM we are using in our investigations. 

%{\color{orange} (Laura, Juergen, Yannick: can we compare with other papers?)}

\section{Summary}

In this paper we have studied compact objects that are made of OM admixed with non-self annihilating DM. We have analysed their masses and radius for different DM particle masses for the weakly and strongly interacting DM. We confirm our previous finding
of stable configurations of OM and DM with radii smaller than 10 km and masses similar to Earth- or Jupiter-like stellar objects
\cite{Tolos:2015qra}.
Moreover, we have determined the second Love-number of these compact objects in view of the recent first measurements of the tidal deformability from the observation of gravitational waves from neutron star mergers.

We find that the mixture of OM and DM in compact objects leads to the appearance of new mass-radius stable solutions with very distinct second Love numbers. Whereas a compact star containing small amounts of DM is almost indistinguishable from normal neutron stars, a large amount of DM inside neutron stars will lead to compact objects with radii smaller than 10 km and masses similar to Earth- or Jupiter-like stellar objects. The second Love numbers of these dark compact stellar objects (or dark compact planets) will then be very different from the ones of normal neutron stars.
 The tidal deformability of these dark compact stellar objects will make them distinguishable and detectable in the merger of compact objects with future GW detectors.

In addition, it turns out that for a not too small admixture of DM inside neutron stars, the second Love number is significantly different compared to  ordinary neutron stars or strange quark stars without DM. As an example, Love numbers of stars made only of strange quark matter are much higher than those presented here for the same compactness, as shown in Ref.~\cite{Hinderer:2009ca}. Those are closer to the Buchdahl limit for incompressible stars, and therefore strange quark stars will be distinguishable from the dark compact objects shown in the present manuscript. This feature can serve as an experimentally accessible observable for the presence of DM inside neutron stars for a known compactness and tidal deformability of the merging compact star which does not rely on the EoS of OM. We expect that future improved measurements coming from next-generation gravitational interferometers will be able to utilize this observable to assess the DM content of merging compact stars, as emphasized in Refs.~\cite{Sennett:2017etc,Maselli:2017vfi,Ellis:2018bkr,Karkevandi:2021ygv,DiGiovanni:2021ejn}.

\begin{acknowledgments}
 L.T. acknowledges support from Agencia Estatal Consejo Superior de Investigaciones Cient\'ificas project Nr. 202050I008, by PID2019-110165GB-I00  financed by MCIN/AEI/10.13039/501100011033 and  by Generalitat Valenciana under contract PROMETEO/2020/023. This research has also been supported by the EU STRONG-2020 project under the program  H2020-INFRAIA-2018-1 grant agreement no. 824093, by PHAROS COST Action CA16214 and  by the CRC-TR 211 'Strong-interaction matter under extreme conditions'- project Nr. 315477589 - TRR 211.
\end{acknowledgments}

\appendix

\bibliography{tidal_dark_rev}% Produces the bibliography via BibTeX.

\end{document}